\documentclass[journal=jacsat,manuscript=article]{achemso}

\usepackage[version=3]{mhchem} 
 \usepackage{tabularray}
 \usepackage{xcolor}

\author{Bruno Ipaves}
\affiliation{Department of Applied Physics and Center for Computational Engineering and Sciences, State University of Campinas, Campinas, São Paulo, Brazil.}
\email{ipaves@unicamp.br}
\author{Raphael B. de Oliveira}
\affiliation{Department of Applied Physics and Center for Computational Engineering and Sciences, State University of Campinas, Campinas, São Paulo, Brazil.}
\author{Guilherme da Silva Lopes Fabris}
\affiliation{Department of Applied Physics and Center for Computational Engineering and Sciences, State University of Campinas, Campinas, São Paulo, Brazil.}
\author{Matthias Batzill}
\affiliation
{Department of Physics, University of South Florida, Tampa, FL 33620, USA.}
\author{Douglas S. Galvão}
\affiliation
{Department of Applied Physics and Center for Computational Engineering and Sciences, State University of Campinas, Campinas, São Paulo, Brazil.}
\email{galvao@ifi.unicamp.br }

\title[An \textsf{achemso} demo]
  {Unraveling Mn intercalation and diffusion in NbSe$_2$ bilayers through DFTB simulations}

\abbreviations{IR,NMR,UV}

\begin{document}







\begin{abstract}
Understanding transition metal atoms' intercalation and diffusion behavior in two-dimensional (2D) materials is essential for advancing their potential in spintronics and other emerging technologies. In this study, we used density functional tight binding (DFTB) simulations to investigate the atomic-scale mechanisms of manganese (Mn) intercalation into NbSe$_2$ bilayers. Our results show that Mn prefers intercalated and embedded positions rather than surface adsorption, as cohesive energy calculations indicate enhanced stability in these configurations. Nudged elastic band (NEB) calculations revealed an energy barrier of 0.68 eV for the migration of Mn into the interlayer, comparable to other substrates, suggesting accessible diffusion pathways. Molecular dynamics (MD) simulations further demonstrated an intercalation concentration-dependent behavior. Mn atoms initially adsorb on the surface and gradually diffuse inward, resulting in an effective intercalation at higher Mn densities before clustering effects emerge. These results provide helpful insights into the diffusion pathways and stability of Mn atoms within NbSe$_2$ bilayers, consistent with experimental observations and offering a deeper understanding of heteroatom intercalation mechanisms in transition metal dichalcogenides.
\end{abstract}

\section{Introduction}

Since the groundbreaking discovery of graphene in 2004, the field of two-dimensional (2D) materials has expanded rapidly, unveiling a diverse array of novel properties that emerge when materials are reduced to a single or a few atomic layers. These properties — including a high surface-area-to-volume ratio, tunable electronic structures, and enhanced mechanical flexibility — position 2D materials as promising candidates for a wide range of technological applications, such as sensors, energy storage, catalysis, and hydrogen evolution reactions \cite{mahapatra2022synthesis, pramanik2024biotene, joseph2023review, ipaves2025enhancing}.

Among the many families of 2D materials, transition metal dichalcogenides (TMDs) have attracted significant attention due to their exceptional electrical, optical, physical, chemical, and mechanical properties \cite{joseph2023review}. Additionally, TMDs offer remarkable structural versatility, enabling the formation of both lateral heterostructures with other monolayer materials and vertical heterostructures by stacking multiple layers \cite{joseph2023review}.
The ability to manipulate matter at the atomic scale has fueled a scientific and technological revolution, opening pathways to design and engineer nanostructures with tailored physical and chemical functionalities \cite{ipaves2019carbon}. This control is particularly compelling in 2D materials, where introducing foreign atoms or layers can significantly modify the material’s behavior \cite{ipaves2024tuning}.

A recent approach involved synthesizing van der Waals (vdW) heterostructures through the reaction of Bi$_2$Se$_3$ with transition metals like Mn or Cr, forming XBi$_2$Se$_4$ (X = Mn or Cr) layers atop a Bi$_2$Se$_3$ substrate \cite{khatun2024solid}. Such systems, which combine magnetic properties with topological insulator states, can exhibit intriguing phenomena, including the quantum anomalous Hall (QAH) effect, quantized magnetoelectric effect, and Majorana fermions \cite{khatun2024solid}.
Inspired by this approach, experimental studies have demonstrated the feasibility of intercalating Mn guest atoms into NbSe$_2$ layers, enabling the synthesis of ultrathin, atomically engineered films \cite{pathirage2025intercalation}. However, the underlying mechanisms governing inter-atomic interactions and diffusion pathways in these intercalation systems remain incompletely understood \cite{pathirage2025intercalation}. 

Computational investigations can provide valuable insights into the factors influencing atom ordering and diffusion barriers, offering a deeper understanding of these hetero-atom intercalation systems.
Motivated by this context, we systematically explored the interaction between Mn atoms and NbSe$_2$, focusing on the Mn intercalation's structural, energetic, and dynamic aspects to obtain further insights into the microscopic mechanisms that govern this process.

\section{Computational details}
We carried out Density Functional Tight Binding (DFTB) calculations using the DFTB+ code \cite{hourahine2020dftb+} with the Slater-Koster parametrization from the recently developed PTBP dataset \cite{cui2024obtaining}. Structural optimizations were carried out using a rational optimizer, allowing full atomic and lattice relaxation while preserving symmetry. The convergence criteria were set to an energy threshold of $10^{-6}$ Ha, a maximum force threshold of $10^{-3}$ Ha/Bohr, and a self-consistent charge (SCC) convergence threshold of $10^{-3}$ Ha. Along with it, a $\Gamma$-centered $10\times10\times2$ ($10\times10\times1$) k-point grid was employed for 3D (2D) $\rm NbSe_2$ \cite{monkhorst1976special}, and long-range dispersion interactions were included via a Lennard-Jones correction with UFF parameters\cite{zhechkov2005efficient, rappe1992uff}.

The primitive hexagonal unit cell of 3D $\rm NbSe_2$ was created using six atoms \cite{osti_1197331}. For the 2D bilayer, a vacuum region was introduced by setting the lattice parameter perpendicular to the sheets ($z$-axis) to 40 Å, preventing spurious interactions between periodic images.

To understand the stability, the cohesive energy ($E_\text{coh}$) was computed as:
\begin{equation} 
E_\text{coh} = \frac{E_t\big(\rm NbSe_2Mn_x\big) - E_t\big(\rm NbSe_2\big) - xE_t\big({\rm Mn}\big)}{x}, 
\label{eq_E_ads} 
\end{equation}
where $E_t\big(\rm NbSe_2Mn_x\big)$ is the total energy of $\rm NbSe_2$ with $x$ Mn atoms, $E_t\big(\rm NbSe_2\big)$ is the energy of an isolated $\rm NbSe_2$ sheet, and $E_t\big({\rm Mn}\big)$ is the energy of an isolated Mn atom, calculated using the same methodology.

Diffusion barrier values were estimated using the nudged elastic band (NEB) method \cite{henkelman2002methods}, as implemented in the Atomic Simulation Environment (ASE) \cite{ase-paper, ISI:000175131400009}. Two adjacent sites were selected as initial and final reference states, with three interpolated images along the minimum energy pathway (MEP). The improved diameter projection path (IDPP) method was applied to refine the interpolation \cite{smidstrup2014improved}, ensuring a more accurate representation of the transition state. Geometry optimizations were carried out using the BFGS algorithm, with convergence achieved when the maximum force was below 0.05 eV/Å.

Molecular dynamics (MD) simulations were carried out using the velocity Verlet algorithm \cite{verlet1967computer}  with a 1.0 fs integration time step. The system was equilibrated and subsequently propagated for 20 ps at a temperature of 525 K, with a Nosé-Hoover thermostat \cite{evans1985nose}.

\section{Results and discussion}

As mentioned in the introduction, the experimental studies indicate that Mn intercalates into NbSe$_2$ films. However, the exact diffusion mechanism remains unclear \cite{pathirage2025intercalation}. To address this, we systematically investigated the interaction between Mn and $\rm NbSe_2$.

\begin{figure}
\centering
\includegraphics[width=1\linewidth]{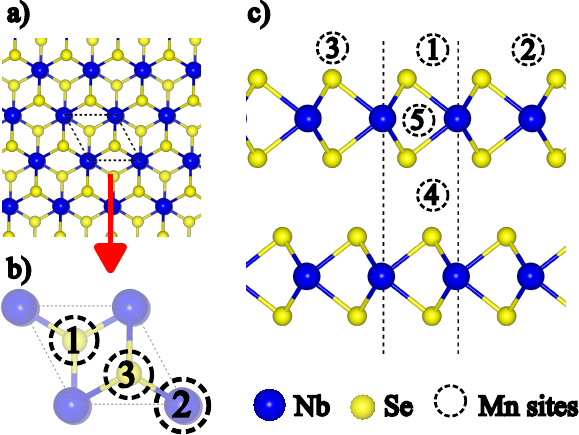}
\caption{(a) Top view, (b) primitive cell, and (c) side view of the $\rm NbSe_2$ structure, showing the possible Mn binding sites: (1) hollow site, (2) top of a Nb atom, (3) top of a Se atom, (4) intercalation between layers, and (5) embedding within a single $\rm NbSe_2$ layer. Each site is indicated with a dashed circle and the corresponding number. The unit cell boundaries are indicated by dashed straight lines and a rhombus. Blue and yellow spheres represent Nb and Se atoms, respectively.}
\label{fig:system}
\end{figure}

Initially, we validate our computational methodology through the structural optimization of bulk (3D) $ \rm NbSe_2$. The optimized lattice parameters, $a = b = 3.608$ Å, and $c = 13.442$ Å, closely match previously reported experimental values \cite{bharucha2024synthesis}. Extending this analysis to a two-dimensional (2D) bilayer, we obtained in-plane lattice parameters of $a = b = 3.607$ Å, showing minimal deviation from the bulk phase, preserving the hexagonal symmetry. To gain deeper insights into Mn interactions with $\rm NbSe_2$, a $4 \times 4$ supercell was created, and its geometry was systematically optimized. The resulting lattice parameters, $a = b = 14.479$ Å, closely matched those of the primitive cell, ensuring a reliable reference for subsequent Mn incorporation. Figure \ref{fig:system} illustrates the investigated systems.

Next, we examined five possible Mn positions, as illustrated in Figure \ref{fig:system} (b-c): three adsorption sites on the surface: (1) hollow site, (2) top of a Nb atom, and (3) top of a Se atom; and two insertion sites within the monolayer: (4) intercalation between layers and (5) embedding within a single $\rm NbSe_2$ layer at the interstitial site. Cohesive energy calculations using Equation \ref{eq_E_ads} revealed that Mn intercalation is more energetically favorable than remaining on the surface, with intercalated Mn ($-5.573$ eV) and embedded Mn ($-5.639$ eV) being more stable than surface-adsorbed Mn ($-3.129$ eV, $-4.483$ eV, and $-5.060$ eV for the Se-top, hollow, and Nb-top sites, respectively).

\begin{figure}
\centering
\includegraphics[width=1\linewidth]{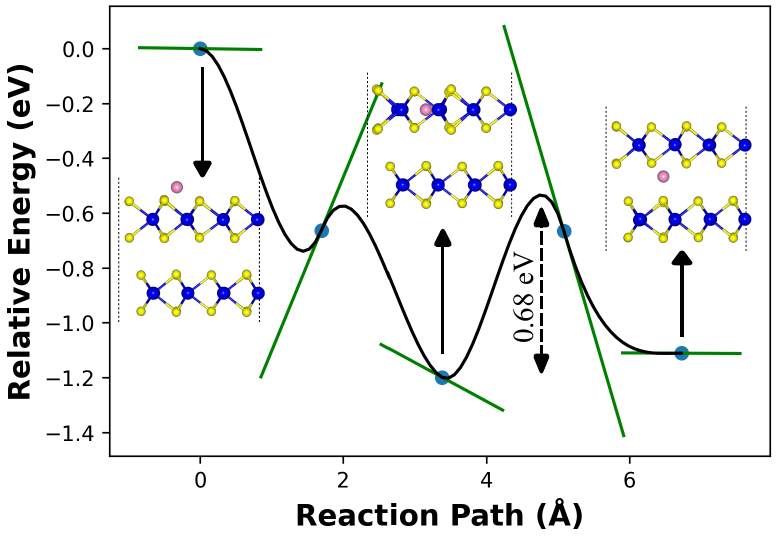}
\caption{Energy profile of Mn diffusion pathways in $\rm NbSe_2$ obtained from nudged elastic band (NEB) calculations. The plot illustrates the energy barriers for Mn migration from the hollow site to the intercalation site between layers, passing through the embedding site within the $\rm NbSe_2$ layer. Blue, yellow, and pink spheres represent Nb, Se, and Nb atoms, respectively.}
\label{fig:neb}
\end{figure}

 Nudged elastic band (NEB) calculations were conducted to clarify Mn migration pathways, as illustrated in Figure \ref{fig:neb}. Our results indicate that Mn can transition from the hollow site to the intercalated position with minimal resistance. Specifically, the energy barrier for the transition from the embedded to the intercalated configuration was determined to be 0.68 eV, a value comparable to those reported for other materials, such as Mn/graphene/Ge (001) and Mn/graphene/GaAs (001) heterostructures (0.1–0.5 eV), and GaAs (0.7–0.8 eV) \cite{edmonds2004mn, strohbeen2021quantifying}. This suggests that Mn can migrate into the van der Waals gap without the need for defects or edge sites, consistent with experimental observations that show a lack of strong dependence on such features for intercalation \cite{pathirage2025intercalation}.

\begin{figure*}
\centering
\includegraphics[width=1\linewidth]{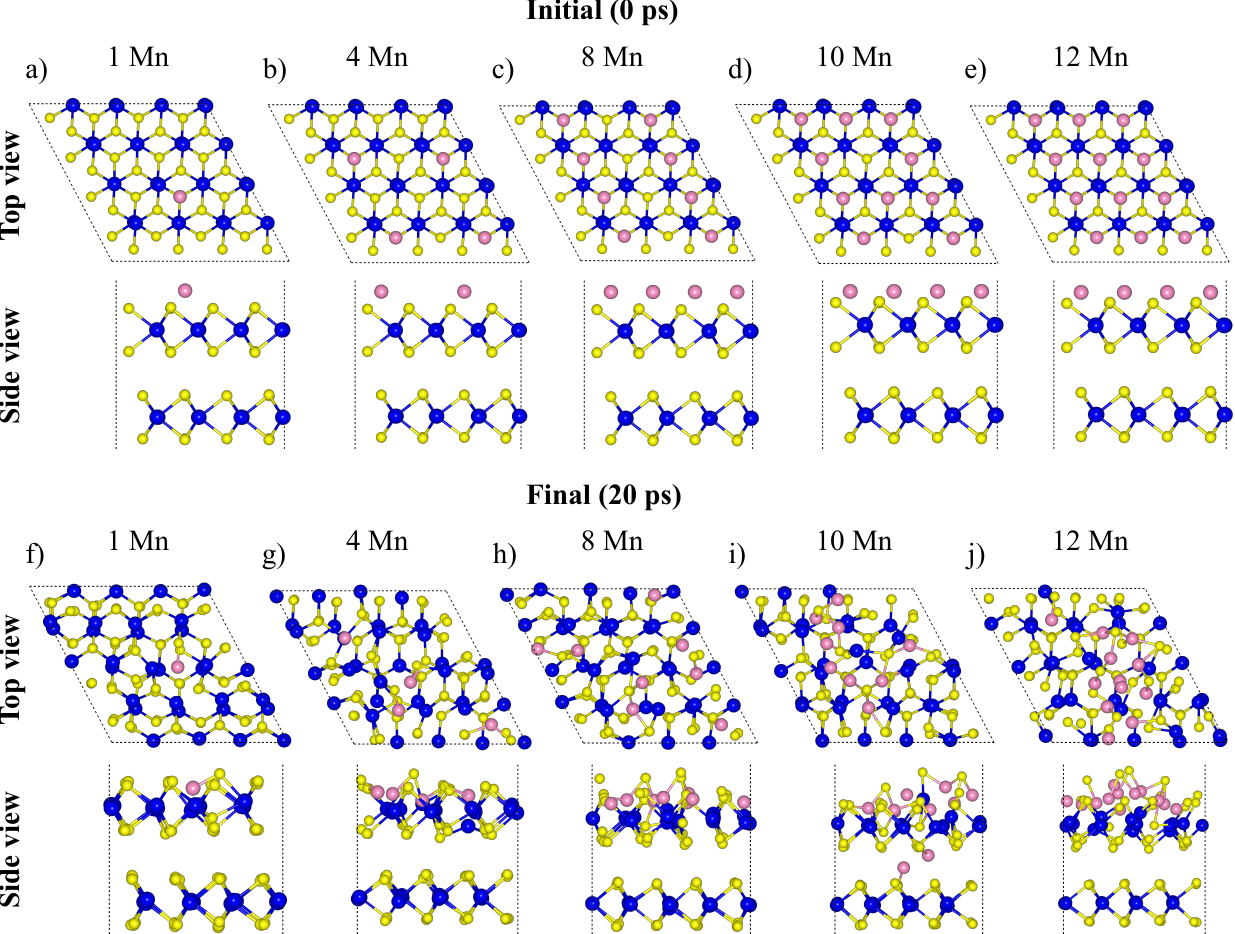}
\caption{Molecular dynamics simulations of Mn intercalation in $\rm NbSe_2$ at varying Mn concentrations. (a)–(e) display top and side views of the initial configurations at 0 ps, while (f)–(j) show the corresponding final configurations at 20 ps. The associated Videos S1–S5 are provided in the Supporting Information. Blue, yellow, and pink spheres represent Nb, Se, and Nb atoms, respectively.}
\label{fig:md}
\end{figure*}

Molecular dynamics (MD) simulations further support this interpretation of intercalation behavior. The simulations were carried out for 20 ps at 525 K, the same temperature used in the experimental studies (250 $^{\circ}$C) \cite{pathirage2025intercalation}. Figure \ref{fig:md} and Videos S1-S5 (Supporting Information) present the MD results for different Mn concentrations (1, 4, 8, 10, and 12 Mn atoms). At lower concentrations (1 Mn atom per supercell), Mn remained on the surface but tended to diffuse inward over time. With 4 and 8 Mn atoms, migration into the interior became more evident, though complete intercalation was not observed. When the concentration increased to 10 Mn atoms, two Mn atoms fully diffused into the layers, confirming the experimental findings of successful intercalation. However, at even higher concentrations (12 Mn atoms), clustering effects emerged, limiting further diffusion. 

These results demonstrate that Mn diffusion to intercalation sites is feasible at temperatures comparable to those in experiments \cite{pathirage2025intercalation}. However, higher Mn concentrations are needed to push the Mn atoms into the intercalation sites compared to the experiments. This may indicate the importance of defects in the experiments, or other intercalation sites with adsorption energies lower than the embedded (interstitial) sites are available, facilitating hopping of Mn into the van der Waals gap.

Our study reveals that Mn embedding within $\rm NbSe_2$ occurs spontaneously under suitable conditions, being energetically more favorable than surface adsorption but less favorable for migration from the embedded site to the van der Waals gap. Additionally, the migration barriers are comparable to those of other systems, indicating that complete Mn diffusion into the van der Waals gap does not require the presence of defects or edge sites. These findings suggest that intercalation could serve as an effective strategy for tuning the electronic and magnetic properties of $\rm NbSe_2$, enabling the design of new functional phases for TMD-based device applications. 

\section{Conclusions}

In this work, density functional tight-binding calculations were employed to investigate Mn intercalation in $\rm NbSe_2$. Our results indicated that Mn embedding within $\rm NbSe_2$ is energetically more favorable than surface adsorption, with low diffusion barriers facilitating spontaneous migration. Furthermore, the migration barrier for complete diffusion is comparable to those in other systems. Molecular dynamics simulations confirmed that intercalation becomes significant at higher Mn concentrations. These findings highlight intercalation as an effective approach for tuning the electronic and magnetic properties of $\rm NbSe_2$, offering potential applications in spintronic devices and the design of novel two-dimensional materials. 

\begin{acknowledgement}
B.I. and R.B.O. thank CNPq process numbers \#153733/2024-1 and \#151043/2024-8, respectively. B.I. and G. S. L. F. thank FAPESP process numbers \#2024/11016-0 and \#2024/03413-9, respectively. D. S. G. acknowledges the Center for Computing in Engineering and Sciences at Unicamp for financial support through the FAPESP/CEPID Grant \#2013/08293-7.
We thank the Coaraci Supercomputer for computer time (Fapesp grant \#2019/17874-0) and the Center for Computing in Engineering and Sciences at Unicamp (Fapesp grant \#2013/08293-4).

\end{acknowledgement}




\bibliography{refs}

\providecommand{\latin}[1]{#1}
\makeatletter
\providecommand{\doi}
  {\begingroup\let\do\@makeother\dospecials
  \catcode`\{=1 \catcode`\}=2 \doi@aux}
\providecommand{\doi@aux}[1]{\endgroup\texttt{#1}}
\makeatother
\providecommand*\mcitethebibliography{\thebibliography}
\csname @ifundefined\endcsname{endmcitethebibliography}  {\let\endmcitethebibliography\endthebibliography}{}
\begin{mcitethebibliography}{24}
\providecommand*\natexlab[1]{#1}
\providecommand*\mciteSetBstSublistMode[1]{}
\providecommand*\mciteSetBstMaxWidthForm[2]{}
\providecommand*\mciteBstWouldAddEndPuncttrue
  {\def\EndOfBibitem{\unskip.}}
\providecommand*\mciteBstWouldAddEndPunctfalse
  {\let\EndOfBibitem\relax}
\providecommand*\mciteSetBstMidEndSepPunct[3]{}
\providecommand*\mciteSetBstSublistLabelBeginEnd[3]{}
\providecommand*\EndOfBibitem{}
\mciteSetBstSublistMode{f}
\mciteSetBstMaxWidthForm{subitem}{(\alph{mcitesubitemcount})}
\mciteSetBstSublistLabelBeginEnd
  {\mcitemaxwidthsubitemform\space}
  {\relax}
  {\relax}

\bibitem[Mahapatra \latin{et~al.}(2022)Mahapatra, Tromer, Pandey, Costin, Lahiri, Chattopadhyay, PM, Roy, Galvao, Kumbhakar, \latin{et~al.} others]{mahapatra2022synthesis}
Mahapatra,~P.~L.; Tromer,~R.; Pandey,~P.; Costin,~G.; Lahiri,~B.; Chattopadhyay,~K.; PM,~A.; Roy,~A.~K.; Galvao,~D.~S.; Kumbhakar,~P.; others Synthesis and characterization of biotene: a new 2D natural oxide from biotite. \emph{Small} \textbf{2022}, \emph{18}, 2201667\relax
\mciteBstWouldAddEndPuncttrue
\mciteSetBstMidEndSepPunct{\mcitedefaultmidpunct}
{\mcitedefaultendpunct}{\mcitedefaultseppunct}\relax
\EndOfBibitem
\bibitem[Pramanik \latin{et~al.}(2024)Pramanik, Mahapatra, Tromer, Xu, Costin, Li, Saju, Alhashim, Pandey, Srivastava, \latin{et~al.} others]{pramanik2024biotene}
Pramanik,~A.; Mahapatra,~P.~L.; Tromer,~R.; Xu,~J.; Costin,~G.; Li,~C.; Saju,~S.; Alhashim,~S.; Pandey,~K.; Srivastava,~A.; others Biotene: earth-Abundant 2D material as sustainable anode for Li/Na-ion battery. \emph{ACS Applied Materials \& Interfaces} \textbf{2024}, \emph{16}, 2417--2427\relax
\mciteBstWouldAddEndPuncttrue
\mciteSetBstMidEndSepPunct{\mcitedefaultmidpunct}
{\mcitedefaultendpunct}{\mcitedefaultseppunct}\relax
\EndOfBibitem
\bibitem[Joseph \latin{et~al.}(2023)Joseph, Mohan, Lakshmy, Thomas, Chakraborty, Thomas, and Kalarikkal]{joseph2023review}
Joseph,~S.; Mohan,~J.; Lakshmy,~S.; Thomas,~S.; Chakraborty,~B.; Thomas,~S.; Kalarikkal,~N. A review of the synthesis, properties, and applications of 2D transition metal dichalcogenides and their heterostructures. \emph{Materials Chemistry and Physics} \textbf{2023}, \emph{297}, 127332\relax
\mciteBstWouldAddEndPuncttrue
\mciteSetBstMidEndSepPunct{\mcitedefaultmidpunct}
{\mcitedefaultendpunct}{\mcitedefaultseppunct}\relax
\EndOfBibitem
\bibitem[Ipaves \latin{et~al.}(2025)Ipaves, Justo, de~Almeida, Assali, and Autreto]{ipaves2025enhancing}
Ipaves,~B.; Justo,~J.~F.; de~Almeida,~J.~M.; Assali,~L.~V.; Autreto,~P. A. d.~S. Enhancing catalyst activity of two-dimensional C $ \_4 $ N $ \_2 $ through doping for the hydrogen evolution reaction. \emph{arXiv preprint arXiv:2502.10863} \textbf{2025}, \relax
\mciteBstWouldAddEndPunctfalse
\mciteSetBstMidEndSepPunct{\mcitedefaultmidpunct}
{}{\mcitedefaultseppunct}\relax
\EndOfBibitem
\bibitem[Ipaves \latin{et~al.}(2019)Ipaves, Justo, and Assali]{ipaves2019carbon}
Ipaves,~B.; Justo,~J.~F.; Assali,~L.~V. Carbon-related bilayers: nanoscale building blocks for self-assembly nanomanufacturing. \emph{The Journal of Physical Chemistry C} \textbf{2019}, \emph{123}, 23195--23204\relax
\mciteBstWouldAddEndPuncttrue
\mciteSetBstMidEndSepPunct{\mcitedefaultmidpunct}
{\mcitedefaultendpunct}{\mcitedefaultseppunct}\relax
\EndOfBibitem
\bibitem[Ipaves \latin{et~al.}(2024)Ipaves, Justo, Sanyal, and Assali]{ipaves2024tuning}
Ipaves,~B.; Justo,~J.~F.; Sanyal,~B.; Assali,~L.~V. Tuning the electronic and mechanical properties of two-dimensional diamond through N and B doping. \emph{ACS Applied Electronic Materials} \textbf{2024}, \emph{6}, 386--393\relax
\mciteBstWouldAddEndPuncttrue
\mciteSetBstMidEndSepPunct{\mcitedefaultmidpunct}
{\mcitedefaultendpunct}{\mcitedefaultseppunct}\relax
\EndOfBibitem
\bibitem[Khatun \latin{et~al.}(2024)Khatun, Alanwoko, Pathirage, de~Oliveira, Tromer, Autreto, Galvao, and Batzill]{khatun2024solid}
Khatun,~S.; Alanwoko,~O.; Pathirage,~V.; de~Oliveira,~C.~C.; Tromer,~R.~M.; Autreto,~P.~A.; Galvao,~D.~S.; Batzill,~M. Solid State Reaction Epitaxy, A New Approach for Synthesizing Van der Waals heterolayers: The Case of Mn and Cr on Bi2Se3. \emph{Advanced Functional Materials} \textbf{2024}, \emph{34}, 2315112\relax
\mciteBstWouldAddEndPuncttrue
\mciteSetBstMidEndSepPunct{\mcitedefaultmidpunct}
{\mcitedefaultendpunct}{\mcitedefaultseppunct}\relax
\EndOfBibitem
\bibitem[Pathirage \latin{et~al.}(2025)Pathirage, Khatun, and Batzill]{pathirage2025intercalation}
Pathirage,~V.; Khatun,~S.; Batzill,~M. Intercalation of Mn in a few layers of NbSe2 by molecular beam epitaxy. \emph{Surface Science} \textbf{2025}, \emph{754}, 122695\relax
\mciteBstWouldAddEndPuncttrue
\mciteSetBstMidEndSepPunct{\mcitedefaultmidpunct}
{\mcitedefaultendpunct}{\mcitedefaultseppunct}\relax
\EndOfBibitem
\bibitem[Hourahine \latin{et~al.}(2020)Hourahine, Aradi, Blum, Bonafe, Buccheri, Camacho, Cevallos, Deshaye, Dumitric{\u{a}}, Dominguez, \latin{et~al.} others]{hourahine2020dftb+}
Hourahine,~B.; Aradi,~B.; Blum,~V.; Bonafe,~F.; Buccheri,~A.; Camacho,~C.; Cevallos,~C.; Deshaye,~M.; Dumitric{\u{a}},~T.; Dominguez,~A.; others DFTB+, a software package for efficient approximate density functional theory based atomistic simulations. \emph{The Journal of chemical physics} \textbf{2020}, \emph{152}\relax
\mciteBstWouldAddEndPuncttrue
\mciteSetBstMidEndSepPunct{\mcitedefaultmidpunct}
{\mcitedefaultendpunct}{\mcitedefaultseppunct}\relax
\EndOfBibitem
\bibitem[Cui \latin{et~al.}(2024)Cui, Reuter, and Margraf]{cui2024obtaining}
Cui,~M.; Reuter,~K.; Margraf,~J.~T. Obtaining robust density functional tight-binding parameters for solids across the periodic table. \emph{Journal of Chemical Theory and Computation} \textbf{2024}, \emph{20}, 5276--5290\relax
\mciteBstWouldAddEndPuncttrue
\mciteSetBstMidEndSepPunct{\mcitedefaultmidpunct}
{\mcitedefaultendpunct}{\mcitedefaultseppunct}\relax
\EndOfBibitem
\bibitem[Monkhorst and Pack(1976)Monkhorst, and Pack]{monkhorst1976special}
Monkhorst,~H.~J.; Pack,~J.~D. Special points for Brillouin-zone integrations. \emph{Physical review B} \textbf{1976}, \emph{13}, 5188\relax
\mciteBstWouldAddEndPuncttrue
\mciteSetBstMidEndSepPunct{\mcitedefaultmidpunct}
{\mcitedefaultendpunct}{\mcitedefaultseppunct}\relax
\EndOfBibitem
\bibitem[Zhechkov \latin{et~al.}(2005)Zhechkov, Heine, Patchkovskii, Seifert, and Duarte]{zhechkov2005efficient}
Zhechkov,~L.; Heine,~T.; Patchkovskii,~S.; Seifert,~G.; Duarte,~H.~A. An efficient a posteriori treatment for dispersion interaction in density-functional-based tight binding. \emph{Journal of Chemical Theory and Computation} \textbf{2005}, \emph{1}, 841--847\relax
\mciteBstWouldAddEndPuncttrue
\mciteSetBstMidEndSepPunct{\mcitedefaultmidpunct}
{\mcitedefaultendpunct}{\mcitedefaultseppunct}\relax
\EndOfBibitem
\bibitem[Rapp{\'e} \latin{et~al.}(1992)Rapp{\'e}, Casewit, Colwell, Goddard~III, and Skiff]{rappe1992uff}
Rapp{\'e},~A.~K.; Casewit,~C.~J.; Colwell,~K.; Goddard~III,~W.~A.; Skiff,~W.~M. UFF, a full periodic table force field for molecular mechanics and molecular dynamics simulations. \emph{Journal of the American chemical society} \textbf{1992}, \emph{114}, 10024--10035\relax
\mciteBstWouldAddEndPuncttrue
\mciteSetBstMidEndSepPunct{\mcitedefaultmidpunct}
{\mcitedefaultendpunct}{\mcitedefaultseppunct}\relax
\EndOfBibitem
\bibitem[Project(2020)]{osti_1197331}
Project,~T.~M. Materials Data on NbSe2 by Materials Project. \textbf{2020}, \relax
\mciteBstWouldAddEndPunctfalse
\mciteSetBstMidEndSepPunct{\mcitedefaultmidpunct}
{}{\mcitedefaultseppunct}\relax
\EndOfBibitem
\bibitem[Henkelman \latin{et~al.}(2002)Henkelman, J{\'o}hannesson, and J{\'o}nsson]{henkelman2002methods}
Henkelman,~G.; J{\'o}hannesson,~G.; J{\'o}nsson,~H. \emph{Theoretical Methods in Condensed Phase Chemistry}; Springer, 2002; pp 269--302\relax
\mciteBstWouldAddEndPuncttrue
\mciteSetBstMidEndSepPunct{\mcitedefaultmidpunct}
{\mcitedefaultendpunct}{\mcitedefaultseppunct}\relax
\EndOfBibitem
\bibitem[Larsen \latin{et~al.}(2017)Larsen, Mortensen, Blomqvist, Castelli, Christensen, Dułak, Friis, Groves, Hammer, Hargus, Hermes, Jennings, Jensen, Kermode, Kitchin, Kolsbjerg, Kubal, Kaasbjerg, Lysgaard, Maronsson, Maxson, Olsen, Pastewka, Peterson, Rostgaard, Schiøtz, Schütt, Strange, Thygesen, Vegge, Vilhelmsen, Walter, Zeng, and Jacobsen]{ase-paper}
Larsen,~A.~H. \latin{et~al.}  The atomic simulation environment—a Python library for working with atoms. \emph{Journal of Physics: Condensed Matter} \textbf{2017}, \emph{29}, 273002\relax
\mciteBstWouldAddEndPuncttrue
\mciteSetBstMidEndSepPunct{\mcitedefaultmidpunct}
{\mcitedefaultendpunct}{\mcitedefaultseppunct}\relax
\EndOfBibitem
\bibitem[Bahn and Jacobsen(2002)Bahn, and Jacobsen]{ISI:000175131400009}
Bahn,~S.~R.; Jacobsen,~K.~W. An object-oriented scripting interface to a legacy electronic structure code. \emph{Comput. Sci. Eng.} \textbf{2002}, \emph{4}, 56--66\relax
\mciteBstWouldAddEndPuncttrue
\mciteSetBstMidEndSepPunct{\mcitedefaultmidpunct}
{\mcitedefaultendpunct}{\mcitedefaultseppunct}\relax
\EndOfBibitem
\bibitem[Smidstrup \latin{et~al.}(2014)Smidstrup, Pedersen, Stokbro, and J{\'o}nsson]{smidstrup2014improved}
Smidstrup,~S.; Pedersen,~A.; Stokbro,~K.; J{\'o}nsson,~H. Improved initial guess for minimum energy path calculations. \emph{The Journal of chemical physics} \textbf{2014}, \emph{140}\relax
\mciteBstWouldAddEndPuncttrue
\mciteSetBstMidEndSepPunct{\mcitedefaultmidpunct}
{\mcitedefaultendpunct}{\mcitedefaultseppunct}\relax
\EndOfBibitem
\bibitem[Verlet(1967)]{verlet1967computer}
Verlet,~L. Computer" experiments" on classical fluids. I. Thermodynamical properties of Lennard-Jones molecules. \emph{Physical review} \textbf{1967}, \emph{159}, 98\relax
\mciteBstWouldAddEndPuncttrue
\mciteSetBstMidEndSepPunct{\mcitedefaultmidpunct}
{\mcitedefaultendpunct}{\mcitedefaultseppunct}\relax
\EndOfBibitem
\bibitem[Evans and Holian(1985)Evans, and Holian]{evans1985nose}
Evans,~D.~J.; Holian,~B.~L. The nose-hoover thermostat. \emph{Journal of Chemical Physics} \textbf{1985}, \emph{83}, 4069--4074\relax
\mciteBstWouldAddEndPuncttrue
\mciteSetBstMidEndSepPunct{\mcitedefaultmidpunct}
{\mcitedefaultendpunct}{\mcitedefaultseppunct}\relax
\EndOfBibitem
\bibitem[Bharucha \latin{et~al.}(2024)Bharucha, Dave, Giri, Chaki, and Limbani]{bharucha2024synthesis}
Bharucha,~S.~R.; Dave,~M.~S.; Giri,~R.~K.; Chaki,~S.~H.; Limbani,~T.~A. Synthesis and mechanistic approach to investigate crystallite size of NbSe2 nanoparticles. \emph{Advances in Natural Sciences: Nanoscience and Nanotechnology} \textbf{2024}, \emph{15}, 015002\relax
\mciteBstWouldAddEndPuncttrue
\mciteSetBstMidEndSepPunct{\mcitedefaultmidpunct}
{\mcitedefaultendpunct}{\mcitedefaultseppunct}\relax
\EndOfBibitem
\bibitem[Edmonds \latin{et~al.}(2004)Edmonds, Bogus{\l}awski, Wang, Campion, Novikov, Farley, Gallagher, Foxon, Sawicki, Dietl, \latin{et~al.} others]{edmonds2004mn}
Edmonds,~K.; Bogus{\l}awski,~P.; Wang,~K.; Campion,~R.~P.; Novikov,~S.; Farley,~N.; Gallagher,~B.; Foxon,~C.; Sawicki,~M.; Dietl,~T.; others Mn Interstitial Diffusion in (G a, M n) A s. \emph{Physical Review Letters} \textbf{2004}, \emph{92}, 037201\relax
\mciteBstWouldAddEndPuncttrue
\mciteSetBstMidEndSepPunct{\mcitedefaultmidpunct}
{\mcitedefaultendpunct}{\mcitedefaultseppunct}\relax
\EndOfBibitem
\bibitem[Strohbeen \latin{et~al.}(2021)Strohbeen, Manzo, Saraswat, Su, Arnold, and Kawasaki]{strohbeen2021quantifying}
Strohbeen,~P.~J.; Manzo,~S.; Saraswat,~V.; Su,~K.; Arnold,~M.~S.; Kawasaki,~J.~K. Quantifying Mn Diffusion through Transferred versus Directly Grown Graphene Barriers. \emph{ACS Applied Materials \& Interfaces} \textbf{2021}, \emph{13}, 42146--42153\relax
\mciteBstWouldAddEndPuncttrue
\mciteSetBstMidEndSepPunct{\mcitedefaultmidpunct}
{\mcitedefaultendpunct}{\mcitedefaultseppunct}\relax
\EndOfBibitem
\end{mcitethebibliography}

\end{document}